\documentclass[aps, prl, twocolumn, superscriptaddress,preprintnumbers]{revtex4-2}
\usepackage{bm, amsmath, amsfonts, amssymb, ascmac, mathtools, braket}
\usepackage{times}
\usepackage{multirow}
\usepackage{graphicx}
\usepackage{float, color, xcolor}

\definecolor{rred}{rgb}{0.8, 0.0, 0.0}
\definecolor{bblue}{rgb}{0.0, 0.0, 0.8}

\usepackage[whole]{bxcjkjatype} 
\usepackage{subfigure} 
\usepackage{amscd} 

\usepackage{bbm}
\usepackage{tabularx}

\usepackage{comment}

\usepackage[
pagebackref=false,
colorlinks=true,
linkcolor=bblue,
urlcolor=bblue,
filecolor=black,
citecolor=rred,
pdfstartview=FitV,
pdftitle={},
pdfauthor={},
pdfsubject={},
pdfkeywords={},
pdfpagemode=None,
bookmarksopen=true
]{hyperref}

\newcommand{\ii}{\text{i}}

\begin{document}

\title{Liouvillian Kramers Degeneracy}

\author{Kohei Kawabata}
\email{kawabata@issp.u-tokyo.ac.jp}
\affiliation{Institute for Solid State Physics, University of Tokyo, Kashiwa, Chiba 277-8581, Japan}

\date{\today}

\begin{abstract}
The Kramers theorem generally guarantees the spectral degeneracy of time-reversal-invariant Hamiltonian operators with half-integer spin,
underlying a variety of physical phenomena.
Despite its fundamental significance, the Kramers degeneracy has been argued to be absent in Liouvillian superoperators acting on mixed states because of the double structure of the Hilbert space.
Here, we instead demonstrate that the interplay of time-reversal symmetry$^{\dag}$ and parity symmetry enforces the Kramers degeneracy in Liouvillians.
We explicitly construct dissipative many-body models of interacting fermions and spins exhibiting the Kramers degeneracy and show that their complex spectral statistics follow those of non-Hermitian random matrices in the corresponding symplectic symmetry classes.
Of the three fundamental universality classes of non-Hermitian random matrices,
the symplectic class has never been realized in symmetry-resolved many-body Liouvillians.
Our constructions provide its explicit realizations and spectral signatures of nonintegrability in symplectic open quantum systems.
\end{abstract}

\maketitle

The Kramers theorem guarantees the degeneracy of energy eigenstates in quantum systems with half-integer spin~\cite{Kramers-30, Wigner-32, Sakurai-textbook, Tasaki-textbook}.
Mathematically, it arises from antiunitary time-reversal symmetry with negative sign,
which defines the symplectic class in the Wigner-Dyson threefold way of Hermitian random matrices~\cite{Wigner-51, *Wigner-58, Dyson-62, Mehta-textbook} and plays a fundamental role in the characterization of quantum chaos~\cite{Berry-Tabor-77, BGS-84, Haake-textbook, Huse-review, Rigol-review}.
The Kramers degeneracy also underlies various condensed matter phenomena,
such as the Anderson localization transitions~\cite{Hikami-80, Ando-89, Kawarabayashi-96, Asada-02, Efetov-textbook, Evers-review} and $\mathbb{Z}_2$ topological insulator phases~\cite{Kane-Mele-05, *Kane-Mele-05-Z2, Roy-09-Z2, *Roy-09, Moore-Balents-07, Fu-Kane-Mele-07, BHZ-06, HK-review, QZ-review, CTSR-review}.

Meanwhile, the physics of open quantum systems has attracted considerable interest.
Unlike isolated quantum systems, open quantum systems interact with their environment,
and the reduced dynamics is no longer governed solely by Hermitian Hamiltonians~\cite{Nielsen-textbook, Breuer-textbook, Rivas-textbook}.
Instead, the open quantum dynamics is generally described by Liouvillian superoperators acting on density matrices.
Understanding such dynamics is important for protecting quantum coherence against dissipation.
By contrast, engineered dissipation can also be useful as a resource for quantum computation and state preparation~\cite{Verstraete-09, Diehl-08}.
In contrast to Hamiltonians, Liouvillians act on operator spaces with a double Hilbert-space structure,
giving rise to a richer symmetry framework for open quantum systems~\cite{Buca-12, Prosen-12, Albert-14, Minganti-18, Lieu-20, McGinley-20, Altland-21, Roberts-21, Kawasaki-22, Kawabata-23}.

The Kramers theorem was extended to generic non-Hermitian operators
under generalized time-reversal symmetry, known as time-reversal symmetry$^{\dag}$, with negative sign~\cite{Sato-11, *Esaki-11, KSUS-19}.
Although many-body Liouvillians are also non-Hermitian superoperators,
their double Hilbert-space structure makes the Kramers degeneracy elusive.
Indeed, whereas the Kramers degeneracy was investigated in dissipative fermionic systems at thermal equilibrium within the Lindblad master equation~\cite{Lieu-22},
Ref.~\cite{Sa-23} subsequently argued that the double Hilbert-space structure of many-body Lindbladians excludes the Kramers degeneracy within individual symmetry-resolved subspaces.
This issue also has important implications for dissipative quantum chaos~\cite{Grobe-88, *Grobe-89, Haake-92, Hamazaki-19, Denisov-19, Can-19JPhysA, Sa-20JPhysA, Akemann-19, Sa-20, Wang-20, JiachenLi-21, Tarnowski-21, GarciaGarcia-22PRX, Prasad-22, Ghosh-22, Kawabata-Ryu-26review, Sa-Ribeiro-Denisov-26review}.
In particular, the complex spectral statistics of nonintegrable open quantum systems are expected to exhibit the universal behavior of non-Hermitian random matrices in the same symmetry classes.
Among the 38 symmetry classes of non-Hermitian random matrices~\cite{Bernard-LeClair-02, KSUS-19},
11 support the Kramers degeneracy,
whereas the remaining 27 do not.
Consequently, only a restricted subset of universality classes can appear in open quantum systems,
if the Kramers degeneracy is indeed forbidden in Liouvillians.

In this Letter, we show how the Kramers degeneracy arises within individual symmetry subspaces of many-body Liouvillians.
We find that the argument of Ref.~\cite{Sa-23} relies on restrictive assumptions about the algebra of symmetry operators and demonstrate that the interplay between time-reversal symmetry$^{\dag}$ and parity symmetry enforces the Liouvillian Kramers degeneracy.
In contrast to Ref.~\cite{Lieu-22}, 
our Liouvillian Kramers degeneracy theorem is independent of particle statistics and applicable to both fermionic and bosonic systems.
To illustrate this, we explicitly construct Lindbladian models of interacting fermions and spins exhibiting the Kramers degeneracy. 
We further show that their complex spectral statistics follow those of non-Hermitian random matrices in the corresponding symmetry classes,
providing characterization of nonintegrability for open quantum systems in the symplectic class.

{\it Kramers theorem}.---Suppose that a generic non-Hermitian operator $H$ respects time-reversal symmetry$^{\dag}$~\cite{KSUS-19}
\begin{equation}
    T H^{\dag} T^{-1} = H,
        \label{eq:TRS-H}
\end{equation}
where $T$ is an antiunitary operator (i.e., $TT^{\dag} = T^{\dag}T = 1$ and $TzT^{-1} = z^*$ for $z \in \mathbb{C}$).
Although this symmetry is equivalent to time-reversal symmetry $THT^{-1} = H$ for Hermitian $H$,
the two symmetries are generally distinct for non-Hermitian $H$~\footnote{Whereas time-reversal symmetry$^{\dag}$ enforces the Kramers degeneracy for arbitrary complex eigenvalues $E \in \mathbb{C}$,
time-reversal symmetry ensures the degeneracy only for real eigenvalues $E \in \mathbb{R}$.}.
Let $E \in \mathbb{C}$ be an eigenvalue of $H$,
and $\ket{\psi}$ and $| \psi \rangle\!\rangle$ be the corresponding right and left eigenvectors, respectively (i.e., $H \ket{\psi} = E \ket{\psi}$, $\langle\!\langle \psi | H = E \langle\!\langle \psi |$).
Equation~\eqref{eq:TRS-H} leads to
\begin{equation}
    H \left( T | \psi \rangle\!\rangle \right) = TH^{\dag} | \psi \rangle\!\rangle = TE^* | \psi \rangle\!\rangle = E \left( T | \psi \rangle\!\rangle \right),
\end{equation}
showing that $T | \psi \rangle\!\rangle$ is also a right eigenvector of $H$ with the same eigenvalue $E$.

Depending on $T^2 = +1$ or $T^2 = -1$,
the relevant symmetry class becomes class AI$^{\dag}$ or AII$^{\dag}$, respectively.
Specifically, $T^2 = -1$ enforces the Kramers degeneracy for arbitrary complex eigenvalues $E \in\mathbb{C}$~\cite{Sato-11, *Esaki-11, KSUS-19}.
Indeed, from $T^{\dag} = -T$, we have
\begin{equation}
    \langle\!\langle \psi | T | \psi \rangle\!\rangle = \langle\!\langle \psi | T^{\dag} | \psi \rangle\!\rangle = - \langle\!\langle \psi | T | \psi \rangle\!\rangle,
\end{equation}
leading to $\langle\!\langle \psi | T | \psi \rangle\!\rangle = 0$.
Accordingly, $T| \psi \rangle\!\rangle$ is biorthogonal to $| \psi \rangle\!\rangle$~\cite{Brody-14} and constitutes an independent eigenvector with the same eigenvalue $E$,
establishing the Kramers degeneracy~\footnote{Alternatively, Eq.~\eqref{eq:TRS-H} with $T^2 = -1$ leads to $\left[ \left( E-H \right) U_T \right]^T = - \left( E-H \right) U_T$ ($T \eqqcolon U_T \mathcal{K}$) and hence $\det \left( E - H \right) = \left( \mathrm{Pf} \left[ \left( E-H \right) U_T \right] \right)^2/\det U_T$, 
implying at least the twofold degeneracy.}.
The Kramers degeneracy underlies various physical phenomena,
such as $\mathbb{Z}_2$ non-Hermitian topology and skin effect~\cite{OKSS-20, *KOS-20, Yi-20, Bergholtz-review, Okuma-Sato-review}.
Furthermore, time-reversal symmetry$^{\dag}$ governs the fundamental threefold universality classes of non-Hermitian random matrices~\cite{Hamazaki-20},
serving as a non-Hermitian counterpart of the Wigner-Dyson threefold way~\cite{Wigner-51, *Wigner-58, Dyson-62}.

In the presence of additional unitary symmetry that commutes with $H$,
such as parity symmetry considered below,
the symmetry classification must be performed separately within each subspace of the corresponding conserved charge.
For example, analyzing the full spectrum without resolving these subspaces obscures the characteristic spectral correlations within individual subspaces,
since the spectra of different subspaces are generally uncorrelated.
For Lindbladians, Ref.~\cite{Sa-23} argued that the double Hilbert-space structure excludes the Kramers degeneracy within individual symmetry subspaces.
By contrast, we below demonstrate the Liouvillian Kramers degeneracy by combining time-reversal symmetry$^{\dag}$ with parity symmetry.

{\it Liouvillian Kramers theorem}.---We consider a Liouvillian superoperator $\mathcal{L}$ acting on operators $X \in \mathcal{B}$,
where $\mathcal{B}$ denotes operator space equipped with the Hilbert-Schmidt inner product.
Similar to Eq.~\eqref{eq:TRS-H}, we impose time-reversal symmetry$^{\dag}$ on $\mathcal{L}$:
\begin{equation}
    \mathcal{T} \mathcal{L}^{\dag} \mathcal{T}^{-1} = \mathcal{L},
        \label{eq:TRS-L}
\end{equation}
where $\mathcal{T}$ is an antiunitary superoperator satisfying 
\begin{equation}
    \mathcal{T}^2 = \left( -1 \right)^{\cal N} \neq 1, \quad \left[ \left( -1 \right)^{\cal N} \right]^2 = 1
        \label{eq:parity-L}
\end{equation}
with a parity superoperator $\left( -1 \right)^{\cal N}$.
If $\mathcal{L}$ is irreducible and possesses no conserved quantities,
$\mathcal{T}^2$ must take a definite sign of either $+1$ or $-1$.
However, since Eqs.~\eqref{eq:TRS-L} and \eqref{eq:parity-L} imply $[\mathcal{L}, \left( -1 \right)^{\cal N}] = 0$,
the entire operator space decomposes into invariant parity subspaces,
for which $\mathcal{T}^2$ takes different signs.
Specifically, in the subspace with $\left( -1 \right)^{\cal N} = -1$, the Liouvillian $\mathcal{L}$ respects time-reversal symmetry$^{\dag}$ with $\mathcal{T}^2 = -1$ and thus exhibits the Kramers degeneracy.

In general, physical Liouvillians $\mathcal{L}$ preserve Hermiticity of operators [i.e., $\left( \mathcal{L} \rho \right)^{\dag} = \mathcal{L} \rho$ for $\rho^{\dag} = \rho \in \mathcal{B}$].
Accordingly, $\mathcal{L}$ respects
\begin{equation}
    \mathcal{J} \mathcal{L} \mathcal{J}^{-1} = \mathcal{L}, \quad \mathcal{J}^2 = +1,
\end{equation}
where $\mathcal{J}$ is antiunitary modular conjugation satisfying $\mathcal{J} X = X^{\dag}$ for $X \in \mathcal{B}$~\cite{Kawabata-23}.
The symmetry classification of $\mathcal{L}$ is determined by the interplay between $\mathcal{T}$ and $\mathcal{J}$, 
as well as whether parity $\left( -1 \right)^{\cal N}$ is strong or weak~\cite{Buca-12, Albert-14}.
Weak parity acts on an operator $X\in \mathcal{B}$ as $\left( -1 \right)^{\cal N} X = \left( -1 \right)^{N} X \left( -1 \right)^{N}$,
with a unitary and Hermitian operator $\left( -1 \right)^N \in \mathcal{B}$.
In the following physical realizations, we take the representation $\mathcal{T} X = TXT^{-1}$ with $T^2 = \pm \left( -1 \right)^N$.
Since $\mathcal{J}$ commutes with $\left( -1 \right)^{\cal N}$, $\mathcal{J}$ is respected within each subspace of $\left( -1 \right)^{\cal N}$.
Consequently, the Liouvillian $\mathcal{L}$ belongs to
\begin{equation}
    \begin{cases}
        \text{class BDI}^{\dag} \left( \mathcal{T}^2 = +1 \right)~\text{for}~\mathcal{B}_{+}, \\
        \text{class DIII}^{\dag} \left( \mathcal{T}^2 = -1 \right)~\text{for}~\mathcal{B}_{-},
    \end{cases} 
        \label{eq:symmetry_class_weak}
\end{equation}
where $\mathcal{B}_{\pm}$ denotes the operator subspace with fixed weak parity $\left( -1 \right)^{\cal N} = \pm 1$, 
and the notation of symmetry classes follows the 38-fold classification of non-Hermitian operators~\cite{KSUS-19}.

On the other hand, when parity is strong symmetry, the left- and right-acting parity superoperators are separately conserved:
$\left( -1 \right)^{{\cal N}_+} X = \left( -1 \right)^{N_+} X$ and $\left( -1 \right)^{{\cal N}_-} X = X \left( -1 \right)^{N_-}$.
The operator space then decomposes into four invariant subspaces $\mathcal{B}_{pq}$ ($p, q = \pm 1$),
labeled by the corresponding strong parity $\left( -1 \right)^{N_+}$ and $\left( -1 \right)^{N_-}$.
Within each subspace,
the time-reversal superoperator $\mathcal{T}$ satisfies $\mathcal{T}^2 = pq$.
Moreover, since modular conjugation $\mathcal{J}$ exchanges $\left( -1 \right)^{N_+}$ and $\left( -1 \right)^{N_-}$,
it preserves individual subspace only for $pq = 1$.
Consequently, the relevant symmetry classes are identified as
\begin{equation}
    \begin{cases}
        \text{class BDI}^{\dag} \left( \mathcal{T}^2 = +1 \right)~\text{for}~\mathcal{B}_{++}, \mathcal{B}_{--}, \\
        \text{class AII}^{\dag} \left( \mathcal{T}^2 = -1 \right)~\text{for}~\mathcal{B}_{+-}, \mathcal{B}_{-+}.
    \end{cases} 
        \label{eq:symmetry_class_strong}
\end{equation}

These constructions circumvent the argument of Ref.~\cite{Sa-23},
which relies on specific assumptions about the representations of symmetry operators.
In our constructions, $T^2 = \pm \left( -1 \right)^N$ yields $\mathcal{T}^2 X = \left( -1 \right)^N X \left( -1 \right)^N$.
Since $T$ preserves each parity subspace of the Hilbert space,
$\mathcal{T}$ acts within $\mathcal{B}_{+-}$ and $\mathcal{B}_{-+}$ separately while squaring to $-1$,
enforcing the Kramers degeneracy even within each subspace.
Notably, class AII$^{\dag}$ in Eq.~\eqref{eq:symmetry_class_strong} and class DIII$^{\dag}$ in Eq.~\eqref{eq:symmetry_class_weak} were previously argued to be absent from Lindbladians~\cite{Sa-23}.
Below, we explicitly demonstrate their realizations in open quantum systems.

\begin{figure*}[t]
\centering
\includegraphics[width=1.0\linewidth]{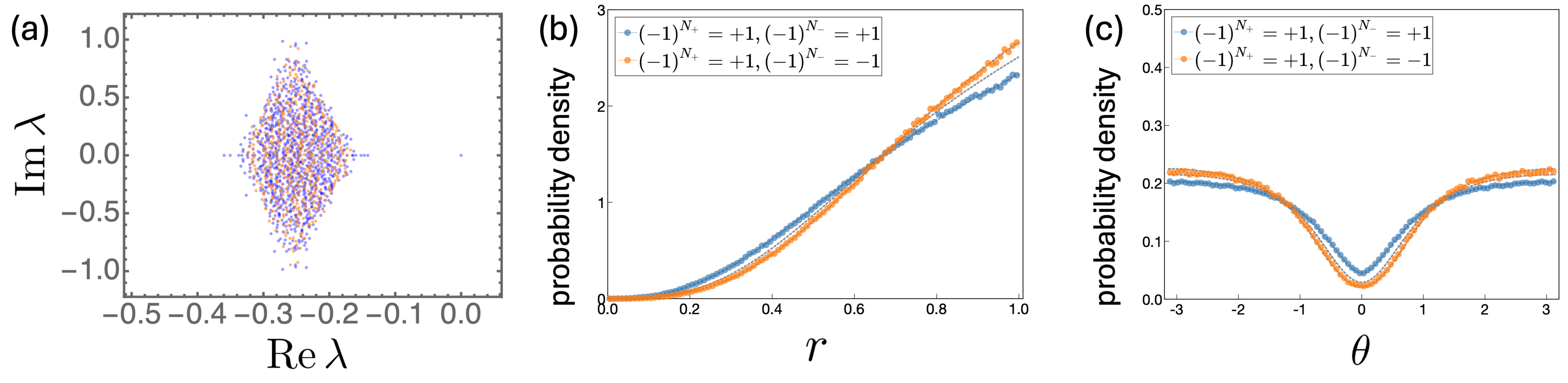} 
\caption{Dissipative interacting fermions for strong parity $\left( -1 \right)^{N_+} = +1$, $\left( -1 \right)^{N_-} = +1$ (blue) and $\left( -1 \right)^{N_+} = +1$, $\left( -1 \right)^{N_-} = -1$ (orange) ($M=3$, $J=1.0$, $\mu = 0.05$, $\gamma_m = 0.25 + 2\varepsilon_m$;
$\varepsilon_m$ is an independent random variable uniformly distributed on $\left[ -0.1, 0.1 \right]$).
(a)~Complex spectrum for a single realization.
(b, c)~Probability densities of the (b)~modulus $r$ and (c)~argument $\theta$ of complex spacing ratios $re^{\ii\theta}$.
$\braket{r} = 0.7220 \pm 0.0002$ and $\braket{\cos \theta} = - 0.1922 \pm 0.0007$ for $\left( -1 \right)^{N_+} = +1$, $\left( -1 \right)^{N_-} = +1$,
and $\braket{r} = 0.7488 \pm 0.0003$ and $\braket{\cos \theta} = - 0.2723 \pm 0.0008$ for $\left( -1 \right)^{N_+} = +1$, $\left( -1 \right)^{N_-} = -1$.
The densities are numerically obtained by averaging over $5000$ independent realizations, 
restricted to the central $50\%$ away from the real axis 
for $\left( -1 \right)^{N_+} = +1$, $\left( -1 \right)^{N_-} = +1$,
and the central $30\%$ including the real axis for $\left( -1 \right)^{N_+} = +1$, $\left( -1 \right)^{N_-} = -1$.
Black dashed curves: non-Hermitian random matrices in classes A, AI$^{\dag}$, and AII$^{\dag}$~\cite{Kawabata-26-CSR}.}	
    \label{fig:SYK}
\end{figure*}

{\it Lindblad master equation}.---While the preceding discussion applies to generic Liouvillians,
we now consider the Kramers degeneracy in the Lindblad master equation $d\rho/dt = \mathcal{L} \rho$, with~\cite{GKS-76, Lindblad-76, Nielsen-textbook, Breuer-textbook, Rivas-textbook}
\begin{equation}
    \mathcal{L} \rho = -\ii \left[ H, \rho \right] + \sum_{n} \left[ L_n \rho L_n^{\dag} - \frac{1}{2} \left\{ L_n^{\dag} L_n, \rho \right\} \right],
\end{equation}
where $H$ is a Hermitian Hamiltonian governing the unitary dynamics, 
and $L_{n}$'s are jump operators describing the dissipative coupling to the environment.
To realize the Lindbladian Kramers degeneracy, we impose time-reversal symmetry on $H$, as in Eq.~\eqref{eq:TRS-H}.
We further require the set of jump operators to be invariant under both time reversal and Hermitian conjugation,
\begin{equation}
    T L_m T^{-1} = \sum_n v_{mn} L_n, \quad L_m^{\dag} = \sum_{n} w_{mn} L_n,
        \label{eq:jump_symmetry}
\end{equation}
where the coefficient matrices $v$ and $w$ are unitary.
A simple realization of these conditions is provided by the set of jump operators, $\{ M, TMT^{-1}, M^{\dag}, TM^{\dag}T^{-1} \}$,
where $M$ denotes an arbitrary operator with definite parity, 
$\left( -1 \right)^{N} M \left( -1 \right)^{N} = \pm M$.
Defining the time-reversal superoperator $\mathcal{T}$ by $\mathcal{T} X \coloneqq TXT^{-1}$ for $X \in \mathcal{B}$,
we find that the Lindbladian $\mathcal{L}$ exhibits the Kramers degeneracy in the weak-parity subspace $\mathcal{T}^2 = \left( -1 \right)^{\cal N} = -1$.
In the following, we provide explicit fermionic and bosonic Lindbladians exhibiting the Kramers degeneracy.
In the End Matter, we also demonstrate its realization in a random Lindbladian (Appendix~\ref{endsec:random}) and an exactly solvable dissipative spinful fermion (Appendix~\ref{endsec:fermion}).

{\it Dissipative interacting fermions}.---Let us consider the Hilbert space consisting of spinful complex fermions $c_{m\sigma}$ ($m = 1, 2, \cdots, M$; $\sigma = \uparrow, \downarrow$). 
Time reversal is implemented by the antiunitary transformation,
\begin{equation}
    Tc_{m\uparrow}T^{-1} = c_{m\downarrow}, \quad Tc_{m\downarrow} T^{-1} = - c_{m\uparrow},
\end{equation}
satisfying
\begin{equation}
    T^2 = \left( -1 \right)^N,\quad N \coloneqq \sum_{m\sigma} n_{m\sigma} = \sum_{m\sigma} c_{m\sigma}^{\dag} c_{m\sigma}.
\end{equation}
We take the Hamiltonian to be 
\begin{equation}
    H = \frac{1}{2} \left( H_{\rm SYK} + T H_{\rm SYK} T^{-1} \right) - \mu N \quad \left( \mu \in \mathbb{R} \right),
\end{equation}
which respects time-reversal symmetry in Eq.~\eqref{eq:TRS-H} by construction.
Here, $H_{\rm SYK}$ denotes the Sachdev-Ye-Kitaev model~\cite{Sachdev-Ye-93, Kitaev-KITP15, Sachdev-15, Polchinski-Rosenhaus-16, Maldacena-Stanford-16}
\begin{equation}
    H_{\rm SYK} = \sum_{1 \leq a < b < c < d \leq 4M} J_{abcd} \psi_a \psi_b \psi_c \psi_d,
\end{equation}
where $\psi_a$'s ($a = 1, 2, \cdots, 4M$) are Majorana fermion operators ($\{ \psi_a, \psi_b \} = \delta_{ab}$, $\psi_a^{\dag} = \psi_a$) constructed from $c_{m\sigma}$,
and the real-valued couplings $J_{abcd} \in \mathbb{R}$ are independently drawn from the Gaussian distribution $\mathcal{N}\,( 0, 3!J^2/\left( 4M \right)^3 )$ ($J \geq 0$).
While $H_{\rm SYK}$ may possess additional symmetry~\cite{You-17, Fu-16, GarciaGarcia-16, Cotler-17},
the chemical potential term $-\mu N \neq 0$ breaks such symmetry.

We introduce random dephasing through the jump operators
\begin{equation}
    L_{m\sigma} = \sqrt{\gamma_m} n_{m\sigma} \quad \left( \gamma_m \geq 0 \right),
\end{equation}
which transform under time reversal and Hermitian conjugation as
\begin{equation}
    TL_{m\uparrow}T^{-1} = L_{m\downarrow},~~TL_{m\downarrow}T^{-1} = L_{m\uparrow};~~L_{m\sigma}^\dag = L_{m\sigma},
\end{equation}
satisfying Eq.~\eqref{eq:jump_symmetry}.
Consequently, the Lindbladian exhibits the Kramers degeneracy in the subspace of weak fermion parity $\left( -1 \right)^{\cal N} = -1$.
Furthermore, these jump operators preserve strong fermion parity [i.e., $\left( -1 \right)^N L_{m\sigma} \left( -1 \right)^N = L_{m\sigma}$],
and the symmetry classification follows Eq.~\eqref{eq:symmetry_class_strong}.

In Fig.~\ref{fig:SYK}\,(a), we show the complex Lindbladian spectra for $\mathcal{B}_{++}$ and $\mathcal{B}_{+-}$.
In our numerical calculations, we sample the dephasing rates $\gamma_m$ independently and uniformly from $[0.05, 0.45]$ for each $m$.
The overall spectra form a characteristic lemon-shaped structure, 
reminiscent of that observed in random Lindbladians~\cite{Denisov-19, Can-19JPhysA}.
Whereas the spectrum in $\mathcal{B}_{++}$ is symmetric about the real axis,
that in $\mathcal{B}_{+-}$ lacks this symmetry,
consistent with the presence or absence of modular conjugation symmetry.
To test the strongly correlated nature of the Lindbladian,
we investigate the distributions of complex spacing ratios $\eta = re^{\ii \theta}$~\cite{Sa-20, Kawabata-26-CSR}.
Given a reference eigenvalue $z \in \mathbb{C}$,
let $z_{\rm NN} \in \mathbb{C}$ and $z_{\rm NNN} \in \mathbb{C}$ denote its nearest- and next-to-nearest-neighbor eigenvalues, respectively.
Then, we define $\eta$ as
\begin{equation}
    \eta \coloneqq \frac{z_{\rm NN} - z}{z_{\rm NNN} - z} \eqqcolon r e^{\ii \theta} \in \mathbb{C},
\end{equation}
whose statistics characterize the local geometry and correlations of the complex spectrum.
In $\mathcal{B}_{+-}$, 
each complex eigenvalue forms a Kramers pair,
which is counted only once for calculations of $\eta$.
In Figs.~\ref{fig:SYK}\,(b, c), we show the probability densities of $r$ and $\theta$,
which agree well with those for non-Hermitian random matrices with time-reversal symmetry$^{\dag}$.
Specifically, the negative weak-fermion-parity subspace $\left( - 1 \right)^{\cal N} = -1$ realizes the symplectic random-matrix statistics coexisting with the Kramers degeneracy,
providing the spectral signatures of nonintegrability for open quantum systems in the symplectic class.

{\it Dissipative spin chain}.---Beyond fermionic systems, our Liouvillian Kramers theorem also applies to bosonic systems.
As an explicit realization, 
we construct a dissipative quantum spin chain exhibiting the Kramers degeneracy.
Let $X_n$, $Y_n$, and $Z_n$ denote Pauli matrices at site $n$ ($n=1, 2, \cdots, L$) of a spin-$1/2$ chain with even length $L \geq 4$.
We choose the Hamiltonian as
\begin{align}
    H &= \sum_{n=1}^{L-1} \left[ J_n^\perp \left( X_n X_{n+1} + Y_n Y_{n+1} \right) + J_n^z Z_n Z_{n+1} \right. \nonumber \\ 
    &\qquad\qquad\qquad\left. + D \left( X_n Y_{n+1} - Y_n X_{n+1} \right)  \right],
\end{align}
where $J_n^{\perp}$ and $J_n^z$ are the bond-dependent transverse and longitudinal exchange interactions, respectively,
and $D$ is the Dzyaloshinskii-Moriya interaction along the $z$ direction.
We further introduce the jump operators
\begin{equation}
    L_n = \sqrt{\gamma} \left[ Z_n + \kappa \left( X_n X_{n+1} - Y_n Y_{n+1} \right) \right]
\end{equation}
for $n=1, 2, \cdots, L-1$, with $\kappa \in \mathbb{R}$.
The two-spin terms in the jumps break conservation of magnetization $\sum_{n} Z_n$ while preserving parity $\prod_{n} Z_n$.

In contrast to the preceding fermionic model,
the relevant antiunitary operation is not conventional spin time reversal but its combination with a uniform $\pi/2$ rotation about the $z$ axis.
Specifically, we introduce an effective antiunitary time-reversal operator
\begin{equation}
    T \coloneqq \left[ \prod_{n=1}^{L} \frac{X_n - Y_n}{\sqrt{2}}\right] \mathcal{K},
\end{equation}
satisfying 
\begin{equation}
    T^2 = \left( -1 \right)^{L/2} \left( -1 \right)^N,\quad \left( -1 \right)^N \coloneqq \prod_{n=1}^{L} Z_n.
\end{equation}
Under this effective time reversal,
the Hamiltonian $H$ is invariant as in Eq.~\eqref{eq:TRS-H}, 
and the jump operators satisfy $T L_n T^{-1} = - L_n$.
Accordingly, the resulting Lindbladian exhibits the Kramers degeneracy in the weak-parity subspace $\left( -1 \right)^{\cal N} = -1$.
Similarly to the fermionic model, the jump operators also preserve strong parity [i.e., $\left( -1 \right)^N L_n \left( -1 \right)^N = L_n$],
leading to the symmetry classification in Eq.~\eqref{eq:symmetry_class_strong}.
Notably, in the discussion following Eq.~(46) of Ref.~\cite{Sa-23}, $\eta_{1, 2}$ defined through $q_{1, 2} = \eta_{1, 2} q^{*}_{1, 2}$ are assumed to be unimodular complex numbers (i.e., proportional to the identity), where $q_{1, 2}$ correspond to the unitary part $U_T$ of $T \coloneqq U_T \mathcal{K}$.
By contrast, $U_T$ in our model satisfies $U_T = \left( -1 \right)^{L/2} \left( -1 \right)^N U_T^*$,
violating the assumption made in Ref.~\cite{Sa-23}.

\begin{figure}[t]
\centering
\includegraphics[width=1.0\linewidth]{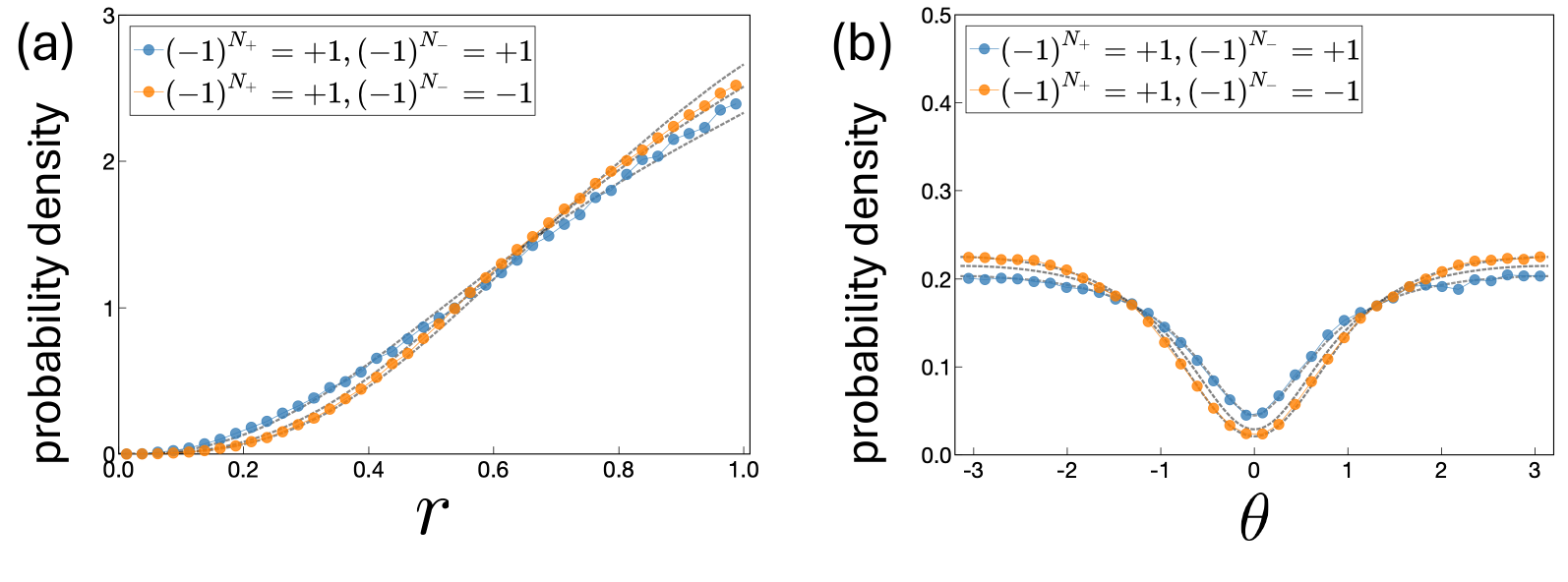} 
\caption{Dissipative quantum spin chain for strong parity $\left( -1 \right)^{N_+} = +1$, $\left( -1 \right)^{N_-} = +1$ (blue) and $\left( -1 \right)^{N_+} = +1$, $\left( -1 \right)^{N_-} = -1$ (orange) [$L=6$, $J_n^{\perp} = 1.0 \left( 1+\varepsilon_n \right)$, $J_n^z = 0.7 \left( 1+\varepsilon'_n \right)$, $D = 0.7$, $\gamma = 1.0$, $\kappa = 0.85$;
$\varepsilon_n$ and $\varepsilon'_n$ are independent random variables uniformly distributed on $\left[ -0.1, 0.1 \right]$].
Probability densities of the (a)~modulus $r$ and (b)~argument $\theta$ of complex spacing ratios $re^{\ii\theta}$.
$\braket{r} = 0.7234 \pm 0.0004$ and $\braket{\cos \theta} = -0.1937 \pm 0.0012$ for $\left( -1 \right)^{N_+} = +1$, $\left( -1 \right)^{N_-} = +1$,
and $\braket{r} = 0.7439 \pm 0.0003$ and $\braket{\cos \theta} = -0.2841 \pm 0.0007$ for $\left( -1 \right)^{N_+} = +1$, $\left( -1 \right)^{N_-} = -1$.
The densities are numerically obtained by averaging over $5000$ independent realizations, 
restricted to the spectral window $\left| \mathrm{Re}\,\lambda + 12.225 \right| < 4$, $4 < \mathrm{Im}\,\lambda < 6$ for $\left( -1 \right)^{N_+} = +1$, $\left( -1 \right)^{N_-} = +1$,
and $\left| \mathrm{Re}\,\lambda + 12.225 \right| < 4$, $\left| \mathrm{Im}\,\lambda \right| < 4$ for $\left( -1 \right)^{N_+} = +1$, $\left( -1 \right)^{N_-} = -1$.
Black dashed curves: non-Hermitian random matrices in classes A, AI$^{\dag}$, and AII$^{\dag}$~\cite{Kawabata-26-CSR}.}	
    \label{fig:spin}
\end{figure}

In Fig.~\ref{fig:spin}, we investigate the complex-spacing-ratio distributions,
focusing on the complex eigenvalues around the spectral center $\mathrm{Tr}\,\mathcal{L}_{pq}/\mathrm{Tr}\,1 = - \gamma\,( 1+2\kappa^2 ) \left( L-1 \right)$.
These probability distributions show good agreement with the random-matrix distributions in the corresponding symmetry classes,
despite deviations near $r=1$ within $\mathcal{B}_{+-}$.
Notably, the dissipative quantum spin chain considered here is spatially local,
and the corresponding Lindbladians $\mathcal{L}_{pq}$ are represented by sparse matrices,
as opposed to dense random matrices.
Nevertheless, the complex spectral statistics of $\mathcal{L}_{pq}$ exhibit the random-matrix universality,
providing a spectral diagnostic of nonintegrability for open quantum systems in the symplectic class.

{\it Discussion}.---We have established the Kramers degeneracy theorem for Liouvillians and demonstrated their random-matrix universality in the symplectic class.
While we have identified classes AII$^{\dag}$ and DIII$^{\dag}$ as the simplest realizations,
additional symmetry can yield a broader range of symplectic classes,
which can be distinguished through the spectral statistics around the symmetry-invariant lines and point~\cite{Xiao-22, *Xiao-24HE}.
As in isolated quantum systems,
the Liouvillian Kramers degeneracy can underlie unique phenomena of open quantum systems.
Although a stationary density matrix can always be chosen in the $\left(-1 \right)^{\cal N} = +1$ subspace,
the $\left(-1 \right)^{\cal N} = -1$ subspace contains no density matrices.
Nevertheless, the Kramers pairs describe independent dynamical modes,
which can be probed through two-time correlations of $\left(-1 \right)^{\cal N} = -1$ operators.
Our results also motivate revisiting the fragility of time-reversal-invariant topological insulators in open quantum systems~\cite{McGinley-20} and exploring possible connections to symmetry fractionalization~\cite{Yagi-25}.

\medskip
\begingroup
\renewcommand{\addcontentsline}[3]{}
\begin{acknowledgments}
K.K. is supported by JSPS KAKENHI Grants 
No.~JP26H02015, No.~JP26K06970, and No.~JP26K17046, 
and JST FOREST Program Grant No.~JPMJFR256P.
\end{acknowledgments}
\endgroup

\let\oldaddcontentsline\addcontentsline
\renewcommand{\addcontentsline}[3]{}
\bibliography{ref.bib}
\let\addcontentsline\oldaddcontentsline

\begin{figure*}[t]
\centering
\includegraphics[width=1.0\linewidth]{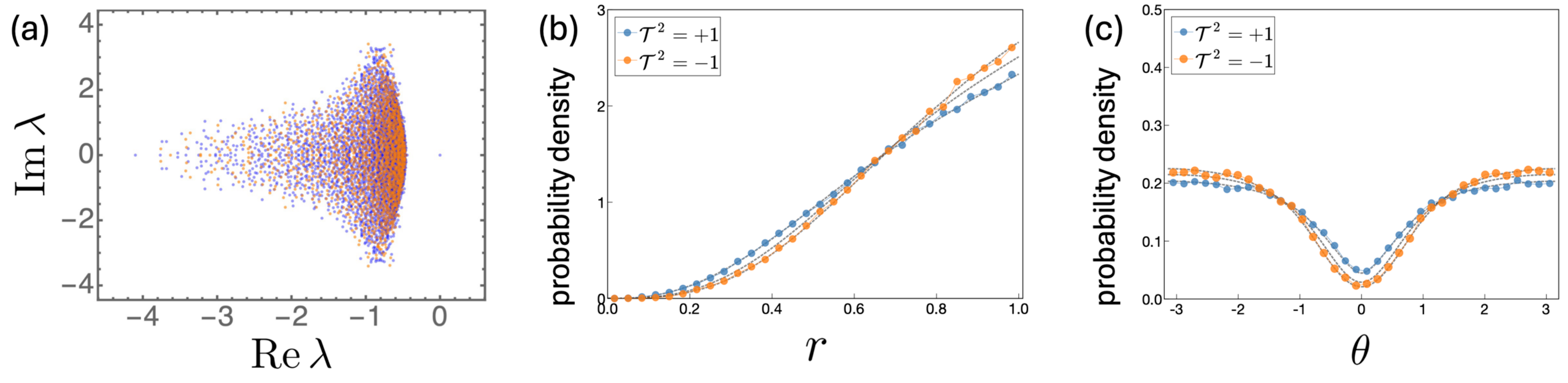} 
\caption{Symplectic random Lindbladian for weak parity $\mathcal{T}^2 = +1$ (blue) and $\mathcal{T}^2 = -1$ (orange) ($D=20$, $h = \gamma = 1$).
(a)~Complex spectrum for a single realization.
(b, c)~Probability densities of the (b)~modulus $r$ and (c)~argument $\theta$ of complex spacing ratios $re^{\ii\theta}$.
$\braket{r} = 0.7228 \pm 0.0014$ and $\braket{\cos \theta} = -0.189 \pm 0.004$ for $\mathcal{T}^2 = +1$,
and $\braket{r} = 0.7487 \pm 0.0018$ and $\braket{\cos \theta} = -0.276 \pm 0.006$ for $\mathcal{T}^2 = -1$.
The densities are numerically obtained by averaging over $1000$ independent realizations, 
restricted to the spectral window $-1.3 < \mathrm{Re}\,\lambda < -0.9$ and $0.8 < \mathrm{Im}\,\lambda < 1.6$.
Black dashed curves: non-Hermitian random matrices in classes A, AI$^{\dag}$, and AII$^{\dag}$~\cite{Kawabata-26-CSR}.}	
    \label{fig:random}
\end{figure*}

\section{End Matter}

\appendix
\setcounter{secnumdepth}{2}

\section{Symplectic random Lindbladian}
    \label{endsec:random}

In a similar manner to Refs.~\cite{Denisov-19, Can-19JPhysA},
we construct a random-matrix model of a Lindbladian exhibiting the Kramers degeneracy.
Let $4D$ ($D \in \mathbb{N}$) be the dimension of the Hilbert space.
We take the Hamiltonian to be
\begin{equation}
    H = \begin{pmatrix}
        H_{\rm GOE} & 0 \\
        0 & H_{\rm GSE}
    \end{pmatrix},
\end{equation}
where $H_{\rm GOE}$ and $H_{\rm GSE}$ are $2D \times 2D$ Hermitian random matrices drawn from the Gaussian orthogonal and symplectic ensembles, respectively.
Specifically, introducing Hermitian random matrices $G_{\pm}$ with the entries
\begin{equation}
    \left( G_{\pm} \right)_{ij} \begin{cases}
        \sim \mathcal{N} \left( 0, 1 \right) & \left( i=j \right), \\
        \sim \mathcal{N}_{\mathbb{C}} \left( 0, 1 \right) & \left( i<j \right), \\
        = \left( G_{\pm} \right)_{ji}^{*} & \left( i>j \right),
    \end{cases}
\end{equation}
we define
\begin{align}
    H_{\rm GOE} &= \frac{h}{\sqrt{2 \left( 2D+1 \right)}} \left( G_+ + G_+^{*} \right), \\
    H_{\rm GSE} &= \frac{h}{\sqrt{2 \left( 2D-1 \right)}} \left( G_- + \Sigma_y G_-^{*} \Sigma_y \right)
\end{align}
with $h \in \mathbb{R}$ and $\Sigma_y \coloneqq\sigma_y \otimes 1_D$.
Here, $\mathcal{N} \left( 0, 1 \right)$ [$\mathcal{N}_{\mathbb{C}} \left( 0, 1\right)$] denotes a real (complex) Gaussian distribution with $\mathbb{E} z = 0$ and $\mathbb{E} \left| z \right|^2 = 1$.
We choose the jump operators as
\begin{equation}
    L_1 = \sqrt{\frac{\gamma}{2}} \begin{pmatrix}
        0 & A \\
        - \Sigma_y A^T & 0
    \end{pmatrix}, \quad L_2 = L_1^{\dag} \quad \left( \gamma \geq 0 \right),
\end{equation}
where $A$ is a non-Hermitian random matrix drawn from the Ginibre unitary ensemble, 
with the independent entries distributed according to $A_{ij} \sim \mathcal{N}_{\mathbb{C}} \left( 0, 1/2D \right)$.

The Hamiltonian $H$ respects time-reversal symmetry in Eq.~\eqref{eq:TRS-H},
implemented by the antiunitary operator
\begin{equation}
    T \coloneqq \begin{pmatrix}
        1_{2D} & 0 \\
        0 & \ii \Sigma_y
    \end{pmatrix} \mathcal{K}.
\end{equation}
Here, $T^2 = \Sigma_z \coloneqq \sigma_z \otimes 1_{2D}$ defines a parity operator.
Moreover, the jump operators transform into each other under time reversal and Hermitian conjugation [Eq.~\eqref{eq:jump_symmetry}]: 
\begin{align}
    T L_1 T^{-1} = \ii L_2, &\quad T L_2 T^{-1} = -\ii L_1; \\
    L_1^{\dag} = L_2, &\quad L_2^{\dag} = L_1.
\end{align}
Consequently, the Liouvillian exhibits the Kramers degeneracy in the subspace of weak parity $\mathcal{T}^2 = -1$.

In Fig.~\ref{fig:random}\,(a),
we show the complex spectrum for a single realization.
In both weak-parity subspaces,
the complex spectrum is symmetric about the real axis,
owing to modular conjugation symmetry.
We also note that we have $\mathbb{E} H_{\rm GOE}^2 = \mathbb{E} H_{\rm GSE}^2 = h^2 1_{2D}$ and $\mathbb{E}\,( L_1^{\dag} L_1 + L_2^{\dag} L_2 ) = \gamma 1_{4D}$.
Accordingly, choosing $h$ and $\gamma$ to be constants of order unity ensures that the Hamiltonian and dissipation contributions have comparable characteristic scales.
In Figs.~\ref{fig:random}\,(b, c), 
we show the complex-spacing-ratio distributions,
which agree with those of non-Hermitian random matrices in class AI$^{\dag}$ (AII$^{\dag}$) for weak parity $+1$ ($-1$).
Reference~\cite{Sa-23} argued that time-reversal symmetry$^{\dag}$ in Eq.~\eqref{eq:TRS-L} with $\mathcal{T}^2 = -1$ always leads to strong symmetry of Lindbladians.
However, our random symplectic Lindbladian exhibits the Kramers degeneracy protected by time-reversal symmetry$^{\dag}$ with $\mathcal{T}^2 = -1$, 
despite the absence of strong symmetry.

\section{Exactly solvable dissipative spinful fermion}
    \label{endsec:fermion}

We illustrate the Lindbladian Kramers degeneracy with a single spinful orbital,
described by fermion operators $c_{\uparrow}$ and $c_{\downarrow}$.
We take the Hamiltonian to be
\begin{equation}
    H = \mu \left( n_{\uparrow} + n_{\downarrow} \right) + U n_{\uparrow} n_{\downarrow} + \Delta\,( c_{\uparrow}^{\dag} c_{\downarrow}^{\dag} + c_{\downarrow} c_{\uparrow} ),
\end{equation}
with $\mu, U, \Delta \in \mathbb{R}$ and $n_{\sigma} \coloneqq c_{\sigma}^{\dag} c_{\sigma}$.
This Hamiltonian respects time-reversal symmetry in Eq.~\eqref{eq:TRS-H} with
\begin{equation}
    T \coloneqq e^{-\ii \pi S^y} \mathcal{K} = e^{- (\pi/2)\,(c_\uparrow^{\dag} c_{\downarrow} - c_\downarrow^\dag c_\uparrow)} \mathcal{K}.
\end{equation}
Although the superconducting pairing term breaks particle-number conservation,
fermion parity remains preserved:
\begin{equation}
    \left( -1 \right)^N H \left( -1 \right)^N = H, \quad \left( -1 \right)^N \coloneqq \left( -1 \right)^{c_\uparrow^\dag c_\uparrow + c_\downarrow^\dag c_\downarrow}.
\end{equation}

We introduce a set of four jump operators $\{ L, \tilde{L}, L^{\dag}, \tilde{L}^\dag\}$ by
\begin{align}
    L &= c_\uparrow + a c_\uparrow^{\dag} + \ii b c_\uparrow n_\downarrow, \\
    \tilde{L} &= TLT^{-1} = c_\downarrow + ac_\downarrow^\dag - \ii bc_\downarrow n_\uparrow
\end{align}
with $a, b \in \mathbb{R}$.
The jump $L$ coherently combines single-fermion loss and gain for up spin, 
with the loss amplitude $1+\ii b n_\downarrow$ depending on the occupation of down spin.
Since these jump operators change fermion parity, 
the resulting Lindbladian only preserves weak fermion parity $\left( -1 \right)^{\cal N}$.
By construction, this set of jump operators is closed under both time reversal and Hermitian conjugation [Eq.~\eqref{eq:jump_symmetry}].
Consequently, the Lindbladian exhibits the Kramers degeneracy in the subspace of weak fermion parity $\left( -1 \right)^{\cal N} = -1$.

Specifically, the characteristic polynomials $\chi_\pm \left( \lambda \right) \coloneqq \det \left( \lambda - \mathcal{L}_\pm \right)$ in the respective weak-fermion-parity subspaces are given as
\onecolumngrid
\begin{align}
&\chi_+(\lambda)
=\lambda(\lambda+\Gamma)^3
\Bigg\{
(\lambda+\Gamma)(\lambda+2\Gamma)
\left[
(\lambda+\Gamma)^2
+(2\mu+U)^2
-4a^2(b^2+4)
+4\Delta^2
\right]
\nonumber\\
&\qquad\qquad\qquad\qquad\qquad\qquad\qquad\qquad\qquad\qquad
-2b^4
\left[
(\lambda+\Gamma)^2
+(2\mu+U)^2
-4a^2(b^2+4)
\right]
\Bigg\}, \\
&\chi_-(\lambda)
=\Bigg\{
(\lambda+\Gamma)^4
+\left[
2\Delta^2
+\frac{(2\mu+U)^2+U^2-\Gamma^2}{2}
-2a^2(b^2+4)
\right](\lambda+\Gamma)^2
\nonumber\\
&\qquad\qquad\qquad+b^2\left[4a^2b^2-U(2\mu+U)\right]
(\lambda+\Gamma)
+\left[
\Delta^2+\mu(\mu+U)
+(a^2-1)^2-b^2-\frac{b^4}{4}
\right]^2
\nonumber\\
&\qquad\qquad\qquad+b^2(b^2+4)
\left[
\left(\mu+\frac{U}{2}\right)^2
+(a^2-1)^2
\right]
+a^2b^2
\left[
32(a^2+1)-b^4-4\mu^2
\right]
-(4a^2+b^2)U^2
\Bigg\}^{2}
\end{align}
\twocolumngrid
with $\Gamma \coloneqq 2\,( 1+a^2 ) + b^2$.
The perfect-square form of $\chi_-$ demonstrates the Kramers degeneracy.

\end{document}